\documentclass[aps,prl,twocolumn,superscriptaddress,groupedaddress,preprintnumbers]{revtex4}
\usepackage{graphicx}  % needed for figures
\usepackage{dcolumn}   % needed for some tables
\usepackage{bm}        % for math
\usepackage{amssymb}   % for math
\usepackage{xcolor}

\usepackage[hidelinks]{hyperref}
\usepackage{amsmath}

\preprint{MIT-CTP/5918}

\begin{document}

\title{Constraining the Resolution Length of Quark-Gluon Plasma\\ 
With New Jet Substructure Measurements}
\author{Arjun Srinivasan Kudinoor}
\email{kudinoor@mit.edu}
\affiliation{%
 Center for Theoretical Physics --- a Leinweber Institute, Massachusetts Institute of Technology, Cambridge, MA 02139
}%

\author{Daniel Pablos}
\email{pablosdaniel@uniovi.es}
\affiliation{
IGFAE, Universidade de Santiago de Compostela, E-15782 Galicia-Spain
}%

\affiliation{Departamento de F\'isica, Universidad de Oviedo, Avda. Federico Garc\'ia Lorca 18, 33007 Oviedo, Spain}
\affiliation{Instituto Universitario de Ciencias y Tecnolog\'ias Espaciales de Asturias (ICTEA), Calle de la Independencia 13, 33004 Oviedo, Spain}

\author{Krishna Rajagopal}
\email{krishna@mit.edu}
\affiliation{%
 Center for Theoretical Physics --- a Leinweber Institute, Massachusetts Institute of Technology, Cambridge, MA 02139
}%

% \date{\today}

\begin{abstract}
We show that recent measurements of substructure-dependent jet suppression constrain the value of the resolution length of the droplets of quark-gluon plasma (QGP) formed in heavy ion collisions. This resolution length, $L_{\rm res}$, is defined such that the medium can only resolve partons within a jet shower that are 
separated by more than $L_{\rm res}$. 
We first use Hybrid Model calculations to reproduce ALICE measurements of 
the scaled Soft Drop angle $\theta_g$
%$R_{\rm AA}$ 
for anti-$k_t$ $R = 0.2$ jets reconstructed from charged-particle tracks. 
%using the Soft Drop grooming procedure with $z_{\rm cut} = 0.2$ and $\beta = 0$. 
We find that the 
narrowing of the $\theta_g$-distribution in PbPb collisions compared to pp collisions that is seen in the ALICE data rules out 
a picture of fully coherent energy loss ($L_{\rm res} = \infty$) where each entire parton shower loses energy to the plasma as if it were a single unresolved colored object. We then use Hybrid Model calculations to reproduce ATLAS measurements of $dR_{12}$, the Soft Drop angle
obtained by applying the Soft Drop grooming procedure to all charged-particle tracks in
$R=1$ jets  reconstructed from $R = 0.2$ skinny subjets. 
%
%and then reclustered using the Soft Drop grooming procedure with $z_{\rm cut} = 0.15$ and $\beta = 0$ on charged-particle tracks associated with each $R = 1$ jet. 
%
Our analysis demonstrates that the ATLAS measurements of $R_{\rm AA}$ for such $R = 1$ jets as a function of $dR_{12}$ 
are inconsistent with a
picture of fully incoherent energy loss ($L_{\rm res} = 0$) in which every splitting in a parton shower is immediately resolved by the plasma. 
We find that our Hybrid Model calculations agree best with the ATLAS measurements if QGP has a finite, nonzero, resolution length 
$L_{\rm res}\sim (1-2)/(\pi T)$.
For the first time, jet substructure measurements are constraining the resolution length of QGP from below, as well as from above.
\end{abstract}

\maketitle

\textit{Introduction.} --- Relativistic heavy ion collisions at RHIC and the LHC have revealed that the high‑temperature phase of QCD, called quark–gluon plasma (QGP), behaves as a nearly perfect, strongly coupled liquid \cite{PHENIX:2004vcz,BRAHMS:2004adc,PHOBOS:2004zne,STAR:2005gfr,Gyulassy:2004zy}. A particularly compelling way to probe the microscopic structure and dynamics of this hot-QCD phase of matter is to analyze heavy‑ion collision events in which high-energy jets are produced. In particular, investigating how the substructure of a jet is modified due to its passage through a droplet of QGP can teach us about the properties and microscopic structure of QGP.

In this Letter, we focus on studying how substructure-dependent suppression of jets in PbPb collisions can teach us about the resolution length $L_{\rm res}$ of QGP. $L_{\rm res}$ is the length scale defined such that if two partons in a jet are separated by a distance less than $L_{\rm res}$, they behave as if they were a single colored object transferring energy and momentum to the QGP coherently; if and only if they are separated by a distance greater than $L_{\rm res}$ will they interact independently with the plasma.

In either a strongly or weakly coupled gauge theory plasma, the resolution length $L_{\rm res}$ can be understood as arising from the physics of color screening, making it reasonable to expect that it is comparable in magnitude to $\lambda_D$, the Debye screening length between static color charges. 
In a weakly coupled plasma with $N_c=N_f=3$, $\lambda_D^{-2} = (3/2) g^2 T^2$ to lowest order in the gauge coupling $g$, meaning $\lambda_D\simeq 1.25/(\pi T)$ for $\alpha_s=1/3$.
In the strongly coupled plasma of ${\cal N}=4$ SYM theory, $\lambda_D=0.294/(\pi T)$~\cite{Bak:2007fk}, making it reasonable to expect $\lambda_D\sim 1/(\pi T$) in the strongly coupled plasma of QCD~\cite{Hulcher:2017cpt,Casalderrey-Solana:2019ubu}.
In a weakly coupled approach, the resolution $L_{\rm res}$ may also be understood in perturbative QCD in the regime of multiple soft scatterings~\cite{Casalderrey-Solana:2011ule,Mehtar-Tani:2011hma,Armesto:2011ir,Mehtar-Tani:2012mfa,Casalderrey-Solana:2012evi,Dominguez:2019ges,Abreu:2024wka}
or in the regime where a single scattering with momentum transfer $\sim \lambda_D^{-1}$ 
dominates~\cite{Mehtar-Tani:2011lic,Casalderrey-Solana:2015bww}
in terms of a critical coherence angle, 
$\theta_c$, defined such that at angular separations above $\theta_c$, two charges of a color dipole act as independent sources of medium-induced gluon radiation.

%\DP{References of people looking at resolution in substructure observables~\cite{Caucal:2019uvr,Caucal:2021cfb,Pablos:2022mrx,Cunqueiro:2023vxl} also energy correlator papers with $\theta_c$~\cite{Andres:2022ovj,Andres:2023xwr} also Xoan's work on antenna within Improved Opacity Expansion~\cite{Kuzmin:2025fyu}}

%When the parton shower interacts with the medium as it would in a strongly coupled gauge theory plasma, a nonzero resolution length $L_{\rm res}$ can be understood as arising from the physics of color screening in the strongly coupled thermal medium. As was first presented in Ref.~\cite{Hulcher:2017cpt}, the resolution length of QGP must be comparable to, or possibly smaller than, the Debye screening length $\lambda_D$ for color charges in the medium. For a strongly coupled plasma with temperature $T$, $\lambda_D \sim 1/T$. In particular, for a plasma in $\mathcal{N} = 4$ supersymmetric Yang--Mills theory, $\lambda_D \approx 0.3/(\pi T)$~\cite{Bak:2007fk} and $\lambda_D \sim (1–2)/(\pi T)$ in a strongly coupled QGP~\cite{Hulcher:2017cpt}.

Regardless of any prior expectation, constraining the value of 
$L_{\rm res}$ using experimental data is 
of central importance as it is a characteristic, dynamical, property of QGP.
In addition, if $L_{\rm res}$ turns out to be much larger than expected, the QGP droplet would see each jet as if it were a single object. In that case, the study of how passage through QGP changes jet substructure would be much less interesting because only rare large-angle scattering of jet constituents would be sensitive to the microscopic structure of QGP.
% In addition, if $L_{\rm res}$ were to turn out to be significantly larger than expected, this would mean that the QGP sees jets as if they were single objects, which would make the study of how passage through QGP changes jet substructure much less interesting as in that case only rare large-angle scattering of jets would be sensitive to the microscopic structure of QGP. 
If $L_{\rm res}\sim (1-2)/(\pi T)$ as expected, and even more so if it is  smaller, jets with the same total $p_T$ but different substructure will lose energy to the QGP differently,  
with $L_{\rm res}$ dictating how the space‑time evolution of a QCD parton shower changes when it develops inside a droplet of QGP rather than in vacuum. 
There have been a number of previous investigations of 
the consequences for jet shape and substructure observables of varying values of $L_{\rm res}$~\cite{Hulcher:2017cpt,Casalderrey-Solana:2019ubu,Pablos:2022mrx,Kudinoor:2025ilx} or the related $\theta_c$~\cite{Caucal:2019uvr,Caucal:2021cfb,Pablos:2022mrx,Andres:2022ovj,Andres:2023xwr,Cunqueiro:2023vxl,Kuzmin:2025fyu}.
In this Letter, we employ calculations using the hybrid strong/weak coupling model for jet quenching (or simply the Hybrid Model)~\cite{Casalderrey-Solana:2014bpa,Casalderrey-Solana:2015vaa,Casalderrey-Solana:2016jvj,Hulcher:2017cpt,Casalderrey-Solana:2018wrw,Casalderrey-Solana:2019ubu,Hulcher:2022kmn} to show how jet substructure observables recently measured by the ALICE Collaboration~\cite{ALargeIonColliderExperiment:2021mqf} and the ATLAS Collaboration~\cite{ATLAS:2023hso, ATLAS:2025lfb} can be used to constrain the value of $L_{\rm res}$. 
In many model treatments of the interaction between a parton shower and a droplet of QGP, including in the Hybrid Model before the work of Ref.~\cite{Hulcher:2017cpt}, $L_{\rm res}=0$ has been assumed. This means that, for simplicity, each parton in the shower was taken to interact with the QGP independently. This makes finding empirical evidence that $L_{\rm res}$ is nonzero important, so as to quantify the consequences of making the simplifying (but certainly incorrect) assumption that it vanishes.

In the following section, we briefly introduce the Hybrid Model and the implementation of a QGP resolution length within the model. We then compare Hybrid Model calculations with ALICE and ATLAS measurements of jet suppression for different jet samples and as a function of multiple angular observables defined using the Soft Drop grooming procedure~\cite{Larkoski:2014wba}, what we call the \textit{Hard Group} grooming procedure~\cite{ATLAS:2023hso, Kudinoor:2025ilx}, and a combination of these two grooming techniques~\cite{ATLAS:2025lfb}. We demonstrate that Hybrid Model calculations are consistent with all experimental results included in this Letter only if quark-gluon plasma has a finite and nonzero resolution length.

\textit{The Hybrid Model.} --- The hybrid strong/weak coupling model, or simply the Hybrid Model, is a theoretical framework for jet quenching in heavy ion collisions. It is designed to describe the multi-scale processes of jet production and evolution through a strongly coupled plasma, like that which is produced in a heavy ion collision. The Hybrid Model treats the weakly coupled physics of jet production and hard jet evolution perturbatively. Parton splittings that result in a jet shower are determined by the high-$Q^2$, perturbative, DGLAP evolution equations, implemented using PYTHIA 8~\cite{Sjostrand:2007gs,Sjostrand:2014zea}. The soft momentum exchanges between partons in a jet shower and the droplet of QGP through which they propagate dictate that these partons lose energy to the strongly coupled plasma. In the Hybrid Model, each parton in a jet shower loses energy to the plasma as determined by an energy loss formula, calculated holographically in Refs.~\cite{Chesler:2014jva,Chesler:2015nqz} and implemented as described in Refs.~\cite{Casalderrey-Solana:2014bpa, Casalderrey-Solana:2015vaa,Casalderrey-Solana:2016jvj, Hulcher:2017cpt, Casalderrey-Solana:2018wrw,Casalderrey-Solana:2019ubu,Hulcher:2022kmn,Bossi:2024qho, Kudinoor:2025ilx}.
The energy loss formula includes one dimensionless parameter, denoted $\kappa_{\rm sc}$, that governs the strength of the
interaction between the jet parton and the QGP. (For example, in a strongly coupled plasma with temperature $T$ a parton with initial energy $E_{\rm in}$ would lose all of its energy and thermalize over a distance $x_{\rm stop}=E_{\rm in}^{1/3}/(2\kappa_{\rm sc}T^{4/3})$ if it does not split first.) $\kappa_{\rm sc}$ can be calculated in the strongly coupled plasma of ${\cal N}=4$ SYM theory with large 't Hooft coupling $\lambda=g^2 N_c$, where it is given by $1.05\lambda^{1/6}$~\cite{Chesler:2014jva,Chesler:2015nqz}. It is expected to be smaller by a factor $\sim(3-4)$ in QCD, making the stopping distance longer. In the Hybrid Model, we treat $\kappa_{\rm sc}$ as a parameter to be obtained by fitting to data. With $L_{\rm res}=0$ we take $\kappa_{\rm sc}=0.404$ from the fit to measurements of the suppression of jets and high-$p_T$ hadrons in Ref.~\cite{Casalderrey-Solana:2018wrw}. When we employ the Hybrid Model with a nonzero $L_{\rm res}$, we choose the value of $\kappa_{\rm sc}$ so that the suppression of jets with 100-200 GeV matches that obtained in the model with $L_{\rm res}=0$.

Since energy and momentum must be conserved, the momentum and energy lost by a parton is deposited into the plasma, exciting a hydrodynamic wake in the expanding, flowing, and cooling droplet of liquid QGP.
% One can think of the wake sourced by a high energy parton as a portion of the medium that is pulled in the direction of that parton. When the QGP droplet and the wake(s) within it reach the freezeout hypersurface, they hadronize into thousands of soft hadrons, a subset of which are the result of the wake(s) hadronizing at freezeout.
In the Hybrid Model, jet-induced wakes are implemented by generating soft hadrons according to a spectrum determined by employing the Cooper--Frye prescription to the jet-induced perturbation of the stress-energy tensor of the strongly coupled liquid QGP. This spectrum is calculated assuming that the background fluid is longitudinally boost invariant, the jet-induced perturbation to the fluid is small and stays close in rapidity to the rapidity of the jet, and that the jet-induced perturbation to the Cooper-Frye spectrum is small at all 
momenta~\cite{Casalderrey-Solana:2016jvj}. These assumptions, and the shape of the wake-spectrum, are discussed in Refs.~\cite{Casalderrey-Solana:2016jvj,Casalderrey-Solana:2020rsj,Bossi:2024qho,Kudinoor:2025ilx}.

%In addition to jet-induced wakes, there are many other physical effects that can affect the response of the medium to energetic parton showers that traverse it. One such effect is the 
%\textit{QGP resolution length} 

A nonzero QGP resolution length $L_{\rm res}$ was introduced in the Hybrid Model in Ref.~\cite{Hulcher:2017cpt} and is detailed there and in Ref.~\cite{Kudinoor:2025ilx}. 
%is the length scale below which the medium cannot resolve two partons within a jet shower as different sources of energy loss. So, 
Two partons within the same jet shower lose energy independently as a result of their passage through the medium, and deposit energy independently into the medium, if and only if they are separated by a distance larger than $L_{\rm res}$. After a parent parton splits in two, as long as they are separated by a length less than $L_{\rm res}$ they lose energy, and deposit energy into the medium, as if they were still the single parent parton. In this Letter, we shall compare experimental data to  Hybrid Model calculations with four values of $L_{\rm res}$ --- namely 0, $1/(\pi T)$, $2/(\pi T)$, and $\infty$. $L_{\rm res} = 0$ corresponds to fully incoherent energy loss, with every parton in the shower resolved, while $L_{\rm res} = \infty$ corresponds to fully coherent energy loss, with the entire shower losing energy as if it were a single color-charged parton. 
%The implementation of a QGP resolution length within the Hybrid Model is detailed in Refs.~\cite{Hulcher:2017cpt, Kudinoor:2025ilx}.

\textit{Constraining $L_{\rm res}$ with Soft Drop.} --- 
Experimentalists cannot measure the separation between partons in a parton shower as the parton shower develops directly. Rather, they measure the momenta and angle (in azimuth and rapidity) of the hadrons in the final state.
It is reasonable to expect that effects of $L_{\rm res}$ may be discerned by comparing the suppression of jets in which the first (or an early) splitting in the parton shower produces 
daughter partons at larger vs.~smaller angles.
If $L_{\rm res}$ is very large, this splitting angle will not matter as regardless of their separation in angle the daughter partons won't be resolved before they exit the droplet of QGP.  For a smaller value of $L_{\rm res}$, 
daughter partons with a smaller (larger) splitting angle will propagate farther (less far) through the QGP before separating from each other by more than $L_{\rm res}$ and beginning to lose energy independently. Consequently, the corresponding jet will lose less (more) energy.

%A nonzero $L_{\rm res}$ may be understood as being related to a critical coherence angle $\theta_c$ above which two charges of a color dipole act as independent sources of gluon radiation in the regime of multiple soft scatterings in pQCD. A direct experimental handle on $\theta_c$ therefore teaches us about $L_{\mathrm{res}}$, the characteristic distance below which QGP fails to resolve color substructure.

The \textit{Soft Drop} grooming algorithm~\cite{Larkoski:2014wba} is
ideally suited for comparing the energy loss of jets in which an early splitting occurred at larger vs.~smaller angles.
It isolates the first hard branching within a jet by first reconstructing a jet with the anti‑$k_t$ algorithm with radius parameter $R$~\cite{Cacciari:2008gp} and then reclustering its constituents with the Cambridge–Aachen (CA) algorithm \cite{Dokshitzer:1997in, Wobisch:1998wt}, whose ordering follows the angular structure of a parton shower. Stepping through this reclustering history, the algorithm selects the first splitting that fulfills the condition 
\begin{equation} \label{eq:softdrop-condition}
    z>z_{\rm cut}\left(\frac{ R_{12}}{R}\right)^\beta,
\end{equation}
where $\beta$ and $z_{\rm cut}$ are tunable parameters, $z$ is the fraction of transverse momentum carried by the subleading prong in the splitting, and $ R_{12}$ is the angle between the leading and subleading prong. The first hard splitting that satisfies the condition in Eq.~\eqref{eq:softdrop-condition} defines two resulting observables, the groomed momentum fraction $z_g \equiv z$ and the
soft-drop angle $R_g\equiv  R_{12}$, 
where we follow Ref.~\cite{ALargeIonColliderExperiment:2021mqf} and define the
scaled soft-drop angle $\theta_g \equiv R_g/R$. 
%Following the notation from Ref.~\cite{ALargeIonColliderExperiment:2021mqf}, we then define the ratio of angles $\theta_g\equiv R_g/R$, which compares the angle between prongs to the radius parameter used in reconstructing the jet.
We expect that for smaller values of 
$L_{\rm res}$, jets with larger $R_g$ and hence larger $\theta_g$ should lose more energy than jets with smaller $\theta_g$.

%When $\theta_g<\theta_c$, the medium treats the two prongs as a single coherent dipole --- so they lose energy to the plasma as a single source of energy loss. When $\theta_g > \theta_c$ the medium resolves the two prongs as independent color charges that lose energy to the plasma as two distinct sources of energy loss, resulting in higher suppression when compared, on average, to a jet with $\theta_g < \theta_c$. Thus, the effects of the two regimes $\theta_g < \theta_c$ and $\theta_g > \theta_c$ should manifest in measurements of the suppression of jets in heavy ion collisions, as a function of $\theta_g$.

\begin{figure*}[t]
\includegraphics[scale=0.9]{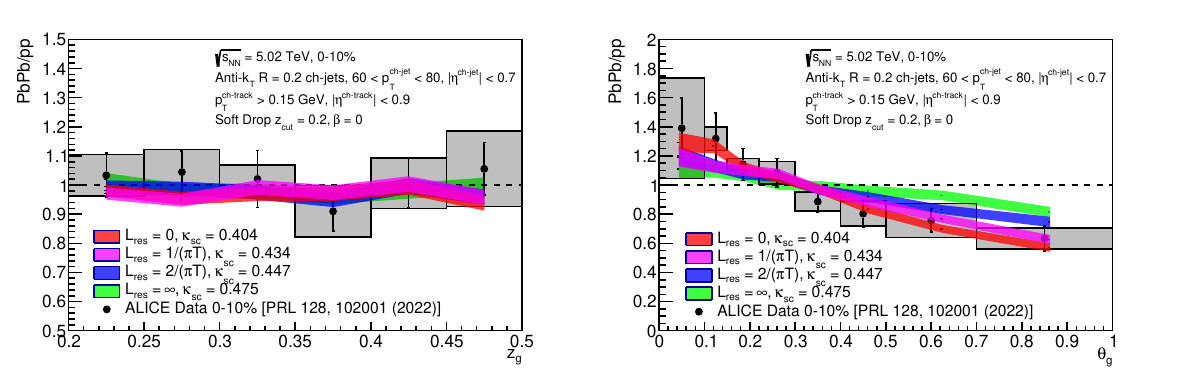}
\caption{
    \label{fig:softdrop}Ratio of the differential cross sections in Eq.~\eqref{eq:diffcross} of jets in PbPb collisions to jets in pp collisions, as a function of $z_g$ (left) and $\theta_g$ (right) for anti-$k_t$ $R = 0.2$ charged-particle jets with $60 < p_T^{\rm ch-jet} < 80$ GeV and $|\eta^{\rm ch-jet}| < 0.7$, reconstructed using the Soft Drop grooming procedure with parameters $z_{\rm cut} = 0.2$ and $\beta = 0$. The colored bands show the results of Hybrid Model calculations with $L_{\rm res} = 0$ (red), $1/(\pi T)$ (purple), $2/(\pi T)$ (blue) and $\infty$ (green). ALICE experimental measurements from Ref.~\cite{ALargeIonColliderExperiment:2021mqf} are depicted using point markers, upon which the vertical bars indicate statistical uncertainties and the shaded boxes indicate systematic uncertainties extracted from Ref.~\cite{ALargeIonColliderExperiment:2021mqf}.
}
\end{figure*}

Fig.~\ref{fig:softdrop} shows the ratios of normalized differential cross sections
\begin{align}\label{eq:diffcross}
    \frac{1}{\sigma_{\text{jets}}}\,\frac{d\sigma}{dz_g} & = \frac{1}{N_{\text{jets}}}\,
    \frac{dN}{dz_g} & \textrm{(left panel)}, \nonumber\\
    \frac{1}{\sigma_{\text{jets}}}\,\frac{d\sigma}{d\theta_g} & = \frac{1}{N_{\text{jets}}}\,
    \frac{dN}{d\theta_g} & \textrm{(right panel)},
\end{align}
of jets in PbPb collisions to jets in pp collisions, as a function of $z_g$ (left panel) and $\theta_g$ (right panel) for anti-$k_t$ $R = 0.2$ charged-jets with $60 < p_T < 80$ GeV and $|\eta| < 0.7$, groomed with $z_{\text{cut}}=0.2$ and $\beta=0$. In Eq.~\eqref{eq:diffcross}, $N$ is the number of jets which pass the Soft Drop condition, $N_{\rm jets}$ is the number of jets which pass the kinematic cuts, and $\sigma$ and $\sigma_{\rm jets}$ are the corresponding cross sections. 
%Our jet reconstruction and selection criteria match those used by ALICE in Ref.~\cite{ALargeIonColliderExperiment:2021mqf}. 
Both panels show the results of Hybrid Model calculations of the PbPb/pp ratio of these differential cross sections upon assuming different values for the QGP resolution length 
$L_{\rm res}$. We compare our results to ALICE data from Ref.~\cite{ALargeIonColliderExperiment:2021mqf}; our jet reconstruction and selection criteria match theirs.
In all our Hybrid Model calculations, the reconstructed jets include soft hadrons originating from jet wakes. We have checked that eliminating these particles changes the plotted  $z_g$ and $\theta_g$ distributions by less than the statistical error bands in Fig.~\ref{fig:softdrop}, confirming that the Soft Drop algorithm grooms away soft hadrons including those originating from jet wakes.

The Hybrid Model curves and the ALICE data in the left panel of Fig.~\ref{fig:softdrop} each show that the PbPb/pp ratio is close to unity, for all $z_g \in (0.2, 0.5)$. This suggests that the momentum fraction shared between the two prongs involved in the first hard splitting of a jet is not modified as jets in a heavy ion collision lose energy while propagating through QGP. %{\color{red}Not clear that this is the only interpretation, so I wrote ``suggests''. -- KR}
%Thus, the differential cross section in Eq.~\ref{eq:diffcross1}, i.e. as a function of $z_g$, should be, and is, unmodified by the medium, for all values of $L_{\rm res}$.

The right panel of Fig.~\ref{fig:softdrop} shows that the differential cross section in Eq.~\eqref{eq:diffcross} as a function of $\theta_g$ {\it is} modified as jets 
propagate through QGP. Let us begin by considering the Hybrid Model calculation with $L_{\rm res}=\infty$, since in that case two jets with the same $p_T$ that have different $\theta_g$ lose the same energy since (by definition) no substructure of jets is resolved.  
This means that the (modest) suppression in the PbPb/pp ratio seen at large $\theta_g$ in the green Hybrid Model calculation must mean that 
jets reconstructed with a given $p_T$ in PbPb collisions have, on average, smaller $\theta_g$
than jets with the same reconstructed $p_T$ in pp collisions.
This can arise for two reasons. First, gluon-initiated jets lose more energy than quark-initiated jets 
even if (as with $L_{\rm res}=\infty$) QGP cannot resolve the differences in how they shower.
This means that a sample of jets with a given $p_T$ after quenching will have a smaller fraction of gluon initiated jets than in vacuum,
meaning a smaller $\theta_g$ on average.
The second reason 
is a consequence of two effects, acting in concert: 
(i) the jets in PbPb collisions have lost energy, on average, which means that imposing the same reconstructed–jet $p_{T}$ selection in PbPb and in pp events preferentially selects higher $p_{T}^{\rm init}$ jet-progenitor partons in the heavy–ion data set; and (ii)
in the collinear approximation used in the soft-drop algorithm~\cite{Larkoski:2014wba}, the angle of a 1-to-2 splitting is
%\begin{equation} \label{eq:one-to-two}
%\theta = \frac{k_T}{z \, p_{T}^{\rm init}},
%\end{equation}
$R_{12}=k_T/zp_T^{\rm init}$
(where $k_T$ is the relative transverse momentum scale of the splitting and $z$ and $p_{T}^{\rm init}$ have already been defined)
which means that at fixed $(k_{T}, z)$, higher-$p_T$ progenitors of jets result in splittings with smaller $ R_{12}$. 
This, together with fact (i) that jets with a given measured $p_T$ began with a higher $p_T^{\rm init}$ in PbPb collisions than in pp collisions, explains the ``tilt'' in the $L_{\rm res}=\infty$ Hybrid Model $\theta_g$ distribution in PbPb collisions relative to pp collisions in the right panel of Fig.~\ref{fig:softdrop}. 

The ``additional tilt'' in the three Hybrid Model curves with finite $L_{\rm res}$, with the PbPb/pp suppression at large $\theta_g$ increasing with decreasing $L_{\rm res}$ in the right panel of Fig.~\ref{fig:softdrop},
arises because (in the Hybrid Model) jets containing multiple resolved sources of energy 
loss lose more energy than an unresolved jet with the same initial $p_T$ and can be understood in two steps. 
First, for any finite $L_{\rm res}$, the larger the value of $\theta_g$ the sooner that
the two subjets from the first hard branching will be resolved. And, the smaller the value of $L_{\rm res}$ the sooner that subjets separated by an angle $R_g=R\theta_g$ will be resolved. So, at any finite value of $L_{\rm res}$ jets lose more energy if their $\theta_g$ is larger, and jets with a given $\theta_g$ lose more energy if $L_{\rm res}$
is smaller. 
We also note that although by its definition $\theta_g$ involves only two subjets, jets with larger $\theta_g$ on average have a larger number of subjets (as demonstrated, for example, in Ref.~\cite{Casalderrey-Solana:2019ubu}) because when the first hard splitting in a jet occurs at a large $\theta_g$ there is more phase space for subsequent splittings. This too means that jets lose more energy if their $\theta_g$ is larger, as long as $L_{\rm res}$ is finite.
%
%\DP{There is also the fact that a higher $\theta_g$ typically involves larger subjet multiplicity, since it means that there is a larger remaining phase-space for splittings down to $\Lambda_{\rm QCD}$. $\theta_g$ is a tagger for jet multiplicity, or number of SoftDrop splittings $n_{SD}$ for example, as shown in the appendix of our paper~\cite{Casalderrey-Solana:2019ubu}}
%
Second,
since the production cross-section for jets is a steeply falling function of $p_T^{\rm init}$, there is a  ``survivor bias'' in PbPb collisions: the jets found in any reconstructed-$p_T$ bin will preferentially be those that lost less energy. This means that jets with larger (smaller) $\theta_g$ will constitute a smaller (larger) fraction of the jets in 
any $p_T$ bin in PbPb collisions than in pp collisions. Consequently, for any finite $L_{\rm res}$ the survivor bias reduces the fraction of jets with larger values of $\theta_g$ in PbPb collisions, and this ``tilt'' becomes stronger for smaller values of $L_{\rm res}$. 
These effects have also been seen in CMS measurements (and Hybrid Model calculations) of
the PbPb/pp ratio for the $\theta_g$ distribution of jets produced in association with a high-$p_T$ photon~\cite{CMS:2024zjn}, where it is possible to reduce (or enhance) the survivor bias by lowering (or raising) $p_T^{\rm jet}/p_T^{\gamma}$
in the jet selection.

From the right panel of Fig.~\ref{fig:softdrop},
we see that the narrowing of the $\theta_g$ distribution in PbPb collisions relative to that in pp collisions seen in the ALICE data is 
inconsistent with our Hybrid Model calculations with $L_{\rm res}=\infty$, in which no substructures within a jet are resolved. The comparison between the ALICE data and our calculations thus indicates that QGP does resolve jet substructure. With the present uncertainties in the measurements, however, these data do not discriminate among the three finite values of $L_{\rm res}$ that we have investigated. 

%\DP{Rewrite} \arjun{See above} However, this kinematic bias does not sufficiently describe the extent to which $\theta_g$ is narrower in PbPb collisions. Even a jet with a given $p_T$ passing through a QGP droplet with $L_{\rm res} = \infty$ will be, on average, more collimated than a jet in vacuum with the same transverse momentum. However, Fig.~\ref{fig:softdrop} shows that the narrowing of $\theta_g$ observed in a plasma with $L_{\rm res} = \infty$ is inconsistent with the ALICE data. Only when one considers a plasma with $L_{\rm res} = 0$, $1/(\pi T)$, or $2/(\pi T)$ do we find good agreement between our Hybrid Model calculations and the experimental measurements. For a plasma with $L_{\rm res} < \infty$, the two prongs in the splitting selected by the Soft Drop algorithm are more likely to be resolved as separate sources of energy loss by the medium when $\theta_g$ is large. Since multiple resolved sources of energy loss within a jet result in higher jet-suppression \cite{ATLAS:2023hso, Kudinoor:2025ilx}, we expect higher suppression of jets with large $\theta_g$ in a plasma with $L_{\rm res} < \infty$, biasing our jet sample towards more energetic progenitors which demonstrate narrower splittings.

\textit{Constraining $L_{\rm res}$ with Hard Group.} --- We note that $\theta_g \sim 1$ for an $R = 0.2$ jet corresponds to an angular separation $R_g = R \,\theta_g \sim 0.2$ between the two prongs selected by the Soft Drop algorithm. From the Hybrid Model results shown in the right panel of Fig.~\ref{fig:softdrop}, it appears that the sensitivity to $L_{\rm res}$ is increasing as one looks at jets with more separated prongs, so it is natural to ask whether we can find methods for studying jets with prongs that are separated by much larger angles.
%If one is able to probe splitting angles beyond the jet radius, then perhaps one can construct observables that are sensitive to the effects of a finite QGP resolution length in the region outside a single jet cone. 
In Refs.~\cite{ATLAS:2023hso, ATLAS:2025lfb}, the ATLAS Collaboration pioneered observables which serve this purpose. In particular, they proposed a method to identify and analyze the substructure of large-radius jets reconstructed from skinny subjets.

In Ref.~\cite{ATLAS:2023hso}, they first reconstructed anti-$k_t$ $R = 0.2$ jets with $|\eta| < 3.0$ and $p_T > 35$ GeV in inclusive jet events. They then use these $R = 0.2$ jets, which we refer to as ``skinny subjets", as the constituents for reconstructing anti-$k_t$ $R = 1.0$ jets with $|y| < 2.0$ and $p_T > 158$~GeV. Finally, the $R = 1.0$ jets were reclustered using the $k_t$-recombination algorithm \cite{Catani:1993hr, Ellis:1993tq} to obtain the observable
\begin{equation}
    \Delta R_{12} \equiv \sqrt{\Delta y_{12}^2 + \Delta \phi_{12}^2},
\end{equation}
defined as the angular separation between the two skinny-subjet-constituents involved in the final reclustering step of the $R = 1.0$ jet. Since the $k_t$ algorithm tends to combine the hardest constituents of a jet last, $\Delta R_{12}$ corresponds to the angular scale of the hardest splitting within a large-radius jet $R = 1.0$ jet. In contrast with the Soft Drop procedure, which moves through the history of a jet's reclustering and \textit{drops soft prongs} at each step of the history, ATLAS' observable~\cite{ATLAS:2023hso} \textit{groups hard prongs} into large-radius jets, which are then reclustered in order to identify the angle of the hardest splitting between skinny-subjet prongs of the large-radius jet. Hence, we call this procedure the \textit{Hard Group} algorithm.
Note also that in the Soft Drop procedure, $R_g$ is the angle between the leading and subleading prongs within the anti-$k_t$ radius of the jet, $R=0.2$ in the ALICE analysis in Fig.~\ref{fig:softdrop}. Contrastingly, in the Hard Group procedure in which a large-radius $R=1.0$ jet is composed from $R=0.2$ skinny subjets, $\Delta R_{12}$ has to be greater than 0.2. And indeed, in their Hard Group analysis~\cite{ATLAS:2023hso}, ATLAS plots results for $0.2<\Delta R_{12}<1.0$.

\begin{figure*}[t]
\includegraphics[scale=0.9]{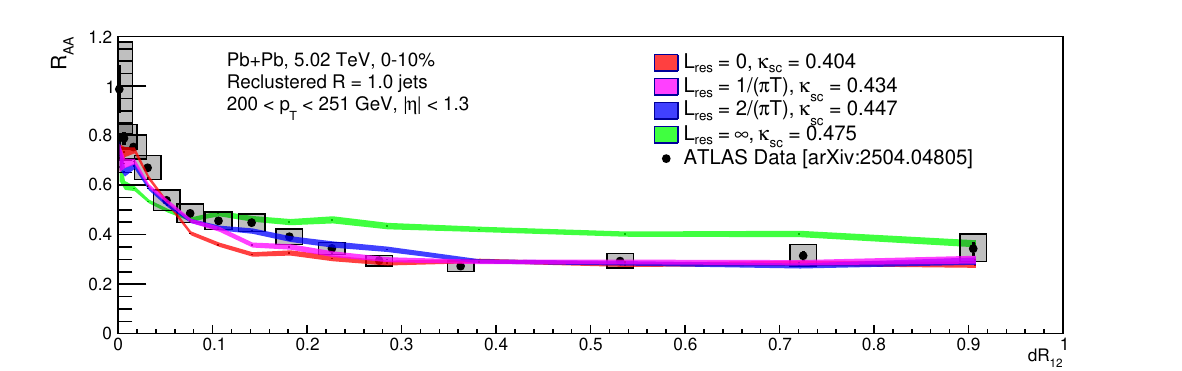}
\caption{
    \label{fig:dr12}$R_{\rm AA}$ as a function of $dR_{12}$ for large-radius $R = 1.0$ jets with $200 < p_T < 251$ GeV and $|\eta| < 1.3$ constructed from skinny subjets using the Hard Group procedure, whose associated charged-particle tracks were then reclustered using the $k_t$ algorithm and the Soft Drop grooming procedure with parameters $z_{\rm cut} = 0.15$ and $\beta = 0$. The colored bands show the results of Hybrid Model calculations with $L_{\rm res} = 0$ (red), $1/(\pi T)$ (purple), $2/(\pi T)$ (blue) and $\infty$ (green). ATLAS experimental measurements from Ref.~\cite{ATLAS:2025lfb} are depicted using point markers, upon which the vertical bars indicate statistical uncertainties and the shaded boxes indicate systematic uncertainties extracted from Ref.~\cite{ATLAS:2025lfb}.
}
\end{figure*}

In Ref.~\cite{Kudinoor:2025ilx} we employed the Hybrid Model to analyze the suppression of large-radius $R = 1.0$ jets constructed using the Hard Group algorithm as a function of their transverse momentum and $\Delta R_{12}$, for three different values of $L_{\rm res}$ --- namely $L_{\rm res} = 0$, $2/(\pi T)$, and $\infty$. We found that the ATLAS data rules out a picture of fully coherent energy loss ($L_{\rm res} = \infty$), in which an entire parton shower loses energy to the medium as a single object. The large-radius jets with skinny-subjet prong separation across the entire range $0.2<\Delta R_{12}<1.0$ are more suppressed in the ATLAS data than in the Hybrid Model calculation with $L_{\rm res}=\infty$, indicating that $L_{\rm res}$ is finite. 
This is consistent with what we have seen 
above in our comparison of Hybrid Model calculations with $L_{\rm res}=\infty$ 
to ALICE measurements of the Soft Drop angle.
%
%above --- namely that the ALICE measurement of $R_{\rm AA}$ as a function of $\theta_g$ rules out $L_{\rm res} = \infty$. 
Furthermore, we found in Ref.~\cite{Kudinoor:2025ilx} that Hybrid Model calculations with
$L_{\rm res} = 0$ and $2/(\pi T)$ both describe
the ATLAS Hard Group measurements across the range $0.2<\Delta R_{12}<1.0$. This suggested~\cite{Kudinoor:2025ilx} that
the most interesting range of $\Delta R_{12}$ (from the perspective of seeking to constrain the value of $L_{\rm res}$) could be around 0.2, which is larger than what is accessible in the ALICE measurements of Fig.~\ref{fig:softdrop} and smaller than what is accessible in the ATLAS measurements of 
Ref.~\cite{ATLAS:2023hso}.

%. In particular, the fact that two skinny $R = 0.2$ subjets must be separated by a minimum angular distance of $\Delta R_{12} \approx 0.2$ limits the angular granularity one can use to constrain $L_{\rm res}$.

\textit{Constraining $L_{\rm res}$ with Soft Drop and Hard Group.} --- The Soft Drop procedure used by ALICE on charged-particle tracks achieved a high degree of granularity for the angular separation of a hard splitting within a jet, but this separation was limited by the size of the (narrow) jet cone. The Hard Group procedure to construct $\Delta R_{12}$ used by ATLAS on calorimetric jets probed large angle splittings within large-radius jets, but was limited in granularity by the radius  of the skinny subjets from which the large-radius jets were composed. In Ref.~\cite{ATLAS:2025lfb}, ATLAS has designed a clever observable $dR_{12}$, which looks like the Soft Drop $R_g$ when $dR_{12}$ is small and like the Hard Group $\Delta R_{12}$ when $dR_{12}$ is large. $dR_{12}$ is constructed via an algorithm that incorporates both the Hard Group and Soft Drop procedures, as follows.

Large-radius anti-$k_t$ $R = 1.0$ jets with $|\eta| < 1.3$ and $p_T > 158$ GeV are reconstructed from anti-$k_t$ skinny $R = 0.2$ subjets with $|\eta| < 3.0$ and $p_T > 35$ GeV via the Hard Group procedure. Instead of reclustering only the skinny subjet components of the large-radius jets, in the algorithm that ATLAS introduced in Ref.~\cite{ATLAS:2025lfb} the next step is to
recluster all charged-particle tracks associated with the large-radius jet.
This could be done in a straightforward fashion in the Hybrid Model; doing so in data is more challenging because of the background. ATLAS identifies the tracks associated with a large-radius jet via a {\it ghost association} procedure that we have followed also, which goes as follows.
First, select all charged-particle tracks with $p_T > 4$ GeV and $|\eta| < 2.5$ in the event containing a large-radius jet and artificially set their transverse momenta to $p_T = 1$ eV. Add these charged-particle \textit{ghost} tracks with this artificially low $p_T$ to a list of inputs, together with the skinny $R = 0.2$ subjets that
were previously identified via the Hard Group procedure. Then, construct an anti-$k_t$ $R=1.0$ jet from this (unusual) list of inputs. Note that this $R=1.0$ jet will include: (i) the skinny subjets as objects; (ii) tracks that are in fact components of the skinny subjets; and (iii) tracks
around the skinny subjets that are part of the $R=1.0$ jet but that are not in a skinny subjet.
Next, discard the skinny subjet objects, and restore the original $p_T$ to each ghost track constituent of the large-radius jet.
The final ghost track constituents, with their original kinematics restored, are now the charged-particle tracks associated with an $R=1.0$ jet. 
Finally, the charged particle tracks with $p_T > 4$ GeV associated with each large-radius jet are then reclustered using the $k_t$ algorithm with $R = 2.5$ and the Soft Drop grooming condition is applied with $z_{\rm cut} = 0.15$ and $\beta = 0$. 
In brief, take all the charged-particle tracks in an $R=1.0$ jet containing skinny subjets identifed via the Hard 
Group procedure, not only those tracks within subjets, and apply the Soft Drop procedure to all these charged particles. Following ATLAS, we denote the Soft Drop angle of the tracks in an $R=1.0$ jet determined via this procedure by $dR_{12}$.

Fig.~\ref{fig:dr12} shows ATLAS measurements~\cite{ATLAS:2025lfb} of $R_{\rm AA}$ as a function of $dR_{12}$, namely the reduction in the number of jets in a given bin of $dR_{12}$ in PbPb collisions relative to that in the corresponding number of pp collisions,
for large-radius $R = 1.0$ jets with $200 < p_T < 251$ GeV, reconstructed using the combined Hard Group $\rightarrow$ Soft Drop procedure described above, compared to Hybrid Model calculations with different values of $L_{\rm res}$. Note that the analysis extends out to values of $dR_{12}\sim 1.0$, as in the straightforward Hard Group analysis of Ref.~\cite{ATLAS:2023hso}, but with this procedure that synthesizes Hard Group and Soft Drop there is now no obstacle to extending the analysis down to very small values of $dR_{12}$. In Supplemental Material, we demonstrate that the dependence of $R_{AA}$ on $dR_{12}$ that ATLAS has measured cannot be explained as resulting solely from changes in the fractions of quark- and gluon-initiated jets in PbPb collisions relative to pp collisions.

In Fig.~\ref{fig:dr12}, our Hybrid Model calculations are compared with ATLAS data extracted from Ref.~\cite{ATLAS:2025lfb}. As expected and as we concluded~\cite{Kudinoor:2025ilx} from the Hard Group measurements in Ref.~\cite{ATLAS:2023hso}, the ATLAS data at large values of $dR_{12}$ rule out $L_{\rm res} = \infty$. 
%For the case of $L_{\rm res} = \infty$, we understand the decrease in $R_{\rm AA}$ at small $dR_{12}$ to be the result of biasing our jet selection towards jets that are inherently more collimated in PbPb collisions than in pp collisions, 
As in Ref.~\cite{Kudinoor:2025ilx}, we can understand the gradual decrease in $R_{\rm AA}$ at large $dR_{12}$ as arising from the fact that some of these $R = 1.0$ jets may contain subjets which originate from different initial high-energy partons (e.g.~from initial state radiation) and which therefore lose energy to the plasma independently.

Strikingly, the ATLAS data also rule out $L_{\rm res} = 0$ --- a picture of fully incoherent energy loss in which every splitting in a parton shower is immediately resolved by the medium
%--- since this results in an over-suppression of jets with narrow substructure. 
--- since with this hypothesis $R_{AA}$ drops much more rapidly with increasing $dR_{12}$ around 0.1 than is seen in the ATLAS data.
Thus, the experimental data favor a QGP resolution length that is both nonzero and finite! 
A resolution length of $L_{\rm res} = 2/(\pi T)$ 
seems to fit the data somewhat better than
a resolution length of $L_{\rm res} = 1/(\pi T)$,
but we leave a full-fledged Bayesian extraction of empirical constraints on both $L_{\rm res}$ and $\kappa_{\rm sc}$ with quantified uncertainties from the measurements of $dR_{12}$ and many other data sets in concert to future work. 

%This should not be taken to mean that the plasma has a resolution length of $2/(\pi T)$, but should instead motivate further precise theoretical calculations of, model analyses using, and experimental measurements to constrain a nonzero $L_{\rm res}$.

% {\color{red}In both of the preceding two paragraphs, I commented some things out that I either did not think was correct as written or thought need not be said. Arjun, if you think what I commented out needs to be restored, it will need to be clarified first, but I did not think that the bits that I commented out were needed.}

\textit{Conclusion and Outlook.} --- In this Letter, we have demonstrated, for the first time, that a consistent description of substructure‑dependent jet suppression across the Soft Drop, Hard Group, and combined $dR_{12}$ observables measured by ALICE and ATLAS favor a nonzero and finite quark-gluon plasma resolution length. Consistent with previous work, an extreme scenario in which an entire shower loses energy coherently as a single object ($L_{\rm res}=\infty$) is decisively ruled out. New here, a picture in which every parton in a shower is resolved independently ($L_{\rm res}=0$) is disfavored by the ATLAS measurements of the $dR_{12}$ dependence of $R_{AA}$ around $dR_{12}\sim (0.1-0.2)$.
%of large-radius jet suppression as a function of $dR_{12}$.

Establishing for the first time that extant experimental measurements disfavor $L_{\rm res}=0$, meaning that they indicate that the QGP resolution length, a fundamental characteristic of the nature of QGP, is both nonzero and finite
and meaning that 
jets interact with the QGP in a way that is neither fully coherent nor fully incoherent,
opens the door to many further investigations.
In heavy ion phenomenology, this discovery has broad implications for parton-level energy loss kernels in any model.  It also motivates investigating how the value of $L_{\rm res}$ influences the pattern of medium response. The observables that we have employed in the present study are, by design, insensitive to the soft hadrons originating from jet wakes so as to allow us to focus attention on the effects of $L_{\rm res}$, but it would be interesting to investigate how changing the value of $L_{\rm res}$ changes other observables that are sensitive to the shapes of jet wakes in various ways~\cite{Casalderrey-Solana:2016jvj,Pablos:2019ngg,Kudinoor:2025ilx,CMS:2025dua,Yang:2025dqu}.
%
%finite nonzero resolution length which requires interpolating smoothly between descriptions of coherent and incoherent energy loss has broad implications for heavy‑ion phenomenology. It modifies parton‑level energy loss kernels, alters the angular pattern of medium response, and reshapes the wakes that jets excite in the plasma. 
%
Our results also motivate future first‑principles calculations of the resolution length, both via pQCD upon assuming weak coupling and via holography upon assuming strong coupling as well as investigations of its evolution in an expanding, non‑uniform, plasma.

ATLAS' measurement of large-radius jet suppression as a function of $dR_{12}$ is the first experimental measurement which favors a nonzero $L_{\rm res}$. It is, however, too soon to say that $L_{\rm res} = 0$ is definitively ruled out. In fact, the comparisons between Hybrid Model calculations with varying $L_{\rm res}$ and a recent CMS measurement of Lund plane observables~\cite{CMS:2025gdw} in PbPb collisions 
appears to
favor $L_{\rm res} = 0$, although we note that none of the Hybrid Model predictions for that measurement 
match the flatness (lack of dependence on the Lund plane angular observable) of the CMS data points. We also stress that we expect these Lund plane observables to be sensitive to elastic scattering between jet partons and medium partons, an effect that has been investigated in Ref.~\cite{Pablos:2024muu} but that has
to date not been incorporated in the Hybrid Model with nonzero $L_{\rm res}$.  This makes doing so a priority for future investigation. 
We have checked that our Hybrid Model calculations of $R_{\rm AA}$ as a function of $dR_{12}$ with $L_{\rm res}=0$ miss the data in the region $dR_{12}\sim 0.05-0.2$ if we include elastic scatterings, just as they do in Fig.~\ref{fig:dr12}. That said, our work adds further motivation for incorporating elastic scatterings into the model with nonzero $L_{\rm res}$.
It will also be important to exploit photon-jet and $Z$-jet collisions, as the statistics available increase in the high-luminosity LHC era, to measure the Soft Drop, Hard Group, and combined Hard Group $\rightarrow$ Soft Drop observables that we have investigated here as well as Lund plane observables in such systems, 
where selecting events based upon the photon/$Z$ momentum can reduce selection biases that partially mask 
coherence and resolution effects. This is only one example of how future high statistics measurements will enable a more precise experimental determination of $L_{\rm res}$.
Finally, as we noted above, a Bayesian extraction of empirical constraints on both $L_{\rm res}$ and $\kappa_{\rm sc}$ with quantified uncertainties from the measurements of $dR_{12}$ and many other observables including the Lund plane observables measured in Ref.~\cite{CMS:2025gdw}, ultimately in photon-jet and $Z$-jet events, is called for.

%Incorporating the effects of other phenomena --- like elastic scatterings between jet partons and quasiparticles in the medium \cite{Hulcher:2022kmn} --- in concert with a finite QGP resolution length is necessary to constrain $L_{\rm res}$ with higher confidence. It will also enable more accurate and precise calculations of jet quenching observables demanded by forthcoming high-statistics experimental datasets. 
%Finally, exploiting photon–jet and $Z$–jet systems, where the photon/$Z$ calibrates the initial momentum of the parton from which a jet originates, to suppress selection biases that partially mask coherence and resolution effects will enable a more precise experimental handle on $L_{\rm res}$.

In summary, the indication from experimental data that QGP behaves as a colored medium with a resolution length that is nonzero and finite marks a significant advance in our understanding of jet quenching and of QGP. This motivates future studies to transform this qualitative insight into a fully quantitative characterization of the nature of deconfined QCD matter.

\textit{Acknowledgments.} --- We thank Carlota Andrés, Jana Bielcikova, Brian Cole, Dominik Karol Derendarz, Jamie Karthein, Jean Du Plessis, Zachary Hulcher, Riccardo Longo, Martin Rybar, Alba Soto-Ontoso and Rachel Steinhorst for useful discussions. KR acknowledges the hospitality of the CERN Theory Department and the Aspen Center for Physics, which is supported by National Science Foundation grant PHY-2210452. Research supported in part by the U.S.~Department of Energy, Office of Science, Office of Nuclear Physics under grant Contract Number DE-SC0011090, by the European Union's Horizon 2020 research and innovation program under the Marie Sk\l odowska-Curie grant agreement No 101155036 (AntScat), by the European Research Council project ERC-2018-ADG-835105 YoctoLHC, by the Spanish Research State Agency under project 
PID2020-119632GB-I00, by Xunta de Galicia (CIGUS Network of Research Centres) and the European Union, and by Unidad de Excelencia Mar\'ia de Maetzu under project CEX2023-001318-M.
ASK is supported by the National Science
Foundation Graduate Research Fellowship Program under Grant No.~2141064. DP is supported by the Ram\'on y
Cajal fellowship RYC2023-044989-I  funded by Spain’s Ministry of Science, Innovation and
Universities.

\bibliography{bibliography}

%\clearpage
\begin{widetext}
\vspace{0.01in}
\section{SUPPLEMENTAL MATERIAL}
\subsection{Quark- and Gluon-Fraction Modifications Alone Cannot Describe ATLAS $dR_{12}$ Measurements}
Some authors have speculated that the dependence of $R_{AA}$ on variables that quantify the angular separation between jet substructures may be understood as resulting solely from changes in the fractions of quark- and gluon-initiated jets in PbPb conditions relative to pp collisions~\cite{Spousta:2015fca,Qiu:2019sfj,Ringer:2019rfk}. 
In this Supplemental Material, we demonstrate via Fig.~\ref{fig:dr12-qgfraction} that such an argument cannot explain the dependence of $R_{AA}$ on $dR_{12}$ that ATLAS has measured.
%a change in the fraction of quark-initiated jets and the fraction of gluon-initiated jets selected in PbPb collisions when compared to those fractions of jets selected in pp collisions is not sufficient to explain the ATLAS measurements of $R_{\rm AA}$ as a function of $dR_{12}$. 

The yield of jets within a given range of transverse momentum $p_T$ and a given range of $dR_{12}$ in pp collisions is proportional to the quantity 
\begin{equation} \label{eq:pp}
    \left. \frac{dN_{\rm jets}}{dp_T \, d(dR_{12})} \right|_{\rm pp}
    = \left. \frac{dN_{\rm quark-initiated~jets}}{dp_T \, d(dR_{12})} \right|_{\rm pp} + \left. \frac{dN_{\rm gluon-initiated~jets}}{dp_T \, d(dR_{12})} \right|_{\rm pp},
\end{equation}
where the label ``quark-initiated" (``gluon-initiated") refers to the yield of jet showers which were initiated by the production of a high energy quark (gluon) in a hard scattering process that subsequently fragmented, for example as described by PYTHIA 8. In the Hybrid Model, we record the information about the kinematics and flavors of the two partons involved in each initial hard scattering from which jets are produced. For a given $R = 1$ jet, we determine which one of the two initial scatterers the jet is closer to in the $(y, \phi)$ plane, using the distance measure $\Delta r = \sqrt{\Delta y^2 + \Delta \phi^2}$. If the closer initial scatterer to the jet is a quark (gluon), then the jet is tagged as a quark-initiated (gluon-initiated) jet.
Similar to Eq.~\eqref{eq:pp}, the jet yield in PbPb collisions is proportional to the quantity
\begin{equation} \label{eq:pbpb}
    \left. \frac{dN_{\rm jets}}{dp_T \, d(dR_{12})} \right|_{\rm PbPb}
    = \left. \frac{dN_{\rm quark-initiated~jets}}{dp_T \, d(dR_{12})} \right|_{\rm PbPb} + \left. \frac{dN_{\rm gluon-initiated~jets}}{dp_T \, d(dR_{12})} \right|_{\rm PbPb}.
\end{equation}
The ratio of the yield of jets in PbPb collisions to the yield of jets in pp collisions defines the nuclear modification factor, $R_{\rm AA}$. Since gluon-initiated jets are more quenched than quark-initiated jets due to the presence of the strongly coupled QGP medium, the fraction of gluon-initiated jets and the fraction of quark-initiated jets which are selected within a given jet $p_T$ bin are modified in heavy ion collisions where a QGP medium is formed. Thus, $R_{\rm AA}$ is sensitive to the medium-modification of the fractions of quark- and gluon-initiated jets.

\begin{figure*}
\includegraphics[scale=0.9]{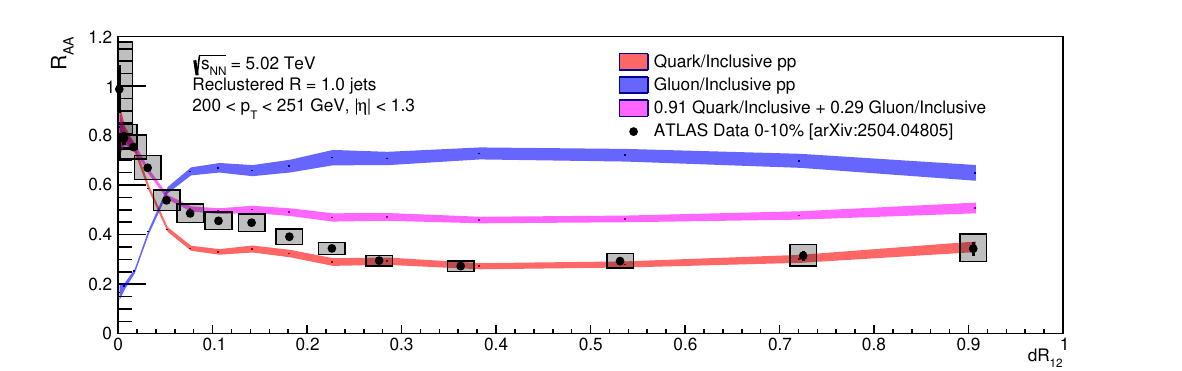}
\caption{
    \label{fig:dr12-qgfraction}Point markers: ATLAS experimental measurements~\cite{ATLAS:2025lfb}, of $R_{\rm AA}$ as a function of $dR_{12}$ for large-radius $R = 1.0$ jets with $200 < p_T < 251$ GeV and $|\eta| < 1.3$ identified via the Hard Group procedure, whose associated charged-particle tracks were then reclustered using the $k_t$ algorithm and the Soft Drop grooming procedure with parameters $z_{\rm cut} = 0.15$ and $\beta = 0$. The vertical bars indicate statistical uncertainties and the shaded boxes indicate systematic uncertainties extracted from Ref.~\cite{ATLAS:2025lfb}. Colored bands: Calculations showing the ratios of the yields of quark-initiated jets (red) and gluon-initiated jets (blue) to inclusive jets in vacuum collisions. 
    The pink band shows the ratio of the jet yield of a sample of jets in vacuum where 91\% of quark-initiated jets and 29\% of gluon-initiated jets are unaltered while the remaining jets are fully quenched and removed from the sample, to an unmodified sample of inclusive jets in vacuum collisions. Note that the calculations shown here are not calculations of $R_{AA}$; they are constructed entirely from calculations of either quark-initiated or gluon-initiated jets in vacuum.
}
\end{figure*}

One may then ask whether experimental measurements of jet suppression, via $R_{\rm AA}$, can be explained {\it solely} by the differing suppression of quark- vs. gluon-initiated jets. As was shown in Ref.~\cite{ALargeIonColliderExperiment:2021mqf}, the ALICE measurement of $R_{\rm AA}$ as a function of the Soft Drop angle $\theta_g$ (that we have discussed above in this Letter) can be explained by a model~\cite{Ringer:2019rfk} in which the only medium effect was the modification of the quark vs. gluon fraction of jet initiators. 
ATLAS has also compared calculations done with this model to their measurements of the $R_g$-dependence 
of $R_{AA}$~\cite{ATLAS:2022vii}, and ALICE has  compared such calculations to their measurements of the angle between two measures of the direction of a jet~\cite{ALICE:2023dwg}. 
However, in Fig.~\ref{fig:dr12-qgfraction} we show that the ATLAS measurements of $R_{\rm AA}$ as a function of $dR_{12}$ across the much wider range of angular scales that this new observable can access cannot be explained solely by the differing suppression of quark- and gluon-initiated jets. Let us unpack this figure and message.

If the only consequence of the propagation of jets through a medium is a change in the fraction of quark- and gluon-initiated jets which are selected, then the PbPb jet yield in Eq.~\eqref{eq:pbpb} is related to the pp jet yield in Eq.~\eqref{eq:pp} by the expressions
\begin{equation}
    \left. \frac{dN_{\rm quark-initiated}}{dp_T \, d(dR_{12})} \right|_{\rm PbPb} = f_q(p_T) \left. \frac{dN_{\rm quark-initiated}}{dp_T \, d(dR_{12})} \right|_{\rm pp}, \quad
    \left. \frac{dN_{\rm gluon-initiated}}{dp_T \, d(dR_{12})} \right|_{\rm PbPb} = f_g(p_T) \left. \frac{dN_{\rm gluon-initiated}}{dp_T \, d(dR_{12})} \right|_{\rm pp},
\end{equation}
where $f_q(p_{T})$ and $f_g(p_T)$ are the fractions of quark-initiated and gluon-initiated jets which survive the jet $p_T$ selection after being quenched by the presence of the QGP medium. Fig.~\ref{fig:dr12-qgfraction} shows $R_{\rm AA}$ for reclustered $R = 1.0$ jets with $200 < p_T < 251$ GeV, as a function of $dR_{12}$ for $(f_q, f_g) =$ $(1, 0)$ (red), $(0, 1)$ (blue), and $(0.91,0.29)$ (pink). $(f_q, f_g) = (1, 0)$ corresponds to the case where all quark jets survive quenching completely unscathed, but no gluon jets survive at all; $(f_q, f_g) = (0, 1)$ corresponds to the case where all gluon jets survive quenching completely unscathed, but no quark jets do at all; $(f_q, f_g) = (0.91, 0.29)$ corresponds to the case where most quark jets survive quenching completely unscathed while some are eliminated, and most gluon jets are eliminated while some survive quenching completely unscathed. The values $f_q = 0.91$ and $f_g = 0.29$ were obtained from Fig. 4 of Ref.~\cite{Qiu:2019sfj}. In particular, we extracted the fraction of quark-initiated (gluon-initiated) jets with $p_{\rm T} \approx 200$ GeV in pp collisions from that figure, finding  0.45 (0.55). We also extracted the number of quark-initiated (gluon-initiated) jets with $p_{\rm T} \approx 200$ GeV in PbPb collisions divided by the total number of jets with the same transverse momentum in pp collisions, finding 0.41 (0.16). Dividing the appropriate quantities yields $f_q(200 {\rm \, GeV}) = 0.41/0.45 \approx 0.91$ and $f_g(200 {\rm \, GeV}) = 0.16/0.55 \approx 0.29$.

It is evident from Fig.~\ref{fig:dr12-qgfraction} that none of these curves can explain the ATLAS experimental data, even qualitatively. 
It is possible to choose a (large) gluon fraction and a (small) quark fraction along the lines of the pink curve in Fig.~\ref{fig:dr12-qgfraction} so as to describe the data at small $dR_{12}$ reasonably well.  Or, it is possible to choose a (quite small) gluon fraction and a (quite large) quark fraction so as to describe the data at large $dR_{12}$. However, no choice of gluon and quark fractions describes the shape of the data across the whole range of $dR_{12}$ over which ATLAS has 
measured $R_{AA}$.
Therefore, only accounting for possible changes in the fraction of quark- and gluon-initiated jets selected in PbPb collisions when compared to pp collisions is insufficient to explain the ATLAS measurements of the $dR_{12}$ dependence of $R_{AA}$.

\clearpage
\end{widetext}

\end{document}